\documentclass[fleqn]{article}
\usepackage{espcrc2}

\usepackage{graphicx}
\usepackage[figuresright]{rotating}

\newcommand{\AmS}{{\protect\the\textfont2
  A\kern-.1667em\lower.5ex\hbox{M}\kern-.125emS}}

\title{Do WMAP data favor neutrino
mass and a coupling between Cold Dark Matter and Dark Energy$\, $?}

\author{G. La Vacca
\address[mib]{Department of Physics G.~Occhialini --
Milano--Bicocca University, Piazza della Scienza 3, 20126 Milano,
Italy } \address[infn]{I.N.F.N., Sezione di Milano},
J.R. Kristiansen\address[ita]{Institute of Theoretical
Astrophysics, University of Oslo, Box 1029, 0315 Oslo, Norway},
L.P.L. Colombo\address[usc]{Department of Physics \& Astronomy,
University of Southern California, Los Angeles, CA 90089-0484},
R. Mainini$\,^{\rm c}$,
S. A. Bonometto$^{\rm a\,\, b}$
}

\begin{document}

%%MISC
\def \aj {AJ}
\def \apj {ApJ}
\def \apjl {ApJL}
\def \mnras {MNRAS}
\def \etal {et~al.~}
\def \eg{e.g.}
\def \Section{\S}
\def \spose#1{\hbox  to 0pt{#1\hss}}  
\def \lta{\mathrel{\spose{\lower 3pt\hbox{$\sim$}}\raise  2.0pt\hbox{$<$}}}
\def \gta{\mathrel{\spose{\lower  3pt\hbox{$\sim$}}\raise 2.0pt\hbox{$>$}}}
%%COSMOLOGY
\def \LCDM {\ifmmode \Lambda{\rm CDM} \else $\Lambda{\rm CDM}$ \fi}
\def \sig8 {\ifmmode \sigma_8 \else $\sigma_8$ \fi} 
\def \OmegaM {\ifmmode \Omega_{\rm M} \else $\Omega_{\rm M}$ \fi} 
\def \OmegaL {\ifmmode \Omega_{\rm \Lambda} \else $\Omega_{\rm \Lambda}$\fi} 
\def \Deltavir {\ifmmode \Delta_{\rm vir} \else $\Delta_{\rm vir}$ \fi}
%DARK MATTER
\def \rs {\ifmmode r_{\rm s} \else $r_{\rm s}$ \fi} 
\def \rrm2 {\ifmmode r_{-2} \else $r_{-2}$ \fi} 
\def \ccm2 {\ifmmode c_{-2} \else$c_{-2}$ \fi} 
\def \cvir {\ifmmode c_{\rm vir} \else $c_{\rm vir}$ \fi} 
\def \cbar {\ifmmode \overline{c} \else $\overline{c}$ \fi}
\def \R200 {\ifmmode R_{200} \else $R_{200}$ \fi} 
\def \Rvir {\ifmmode R_{\rm vir} \else $R_{\rm vir}$ \fi}
\def \v200 {\ifmmode V_{200} \else $V_{200}$ \fi} 
\def \Vvir {\ifmmode V_{\rm  vir} \else  $V_{\rm vir}$  \fi} 
\def  \Vhalo  {\ifmmode V_{\rm halo} \else $V_{\rm halo}$ \fi}
\def \M200 {\ifmmode M_{200} \else $M_{200}$ \fi} 
\def \Mvir {\ifmmode M_{\rm  vir} \else $M_{\rm  vir}$ \fi}  
\def \Mshell  {\ifmmode M_{\rm shell} \else $M_{\rm shell}$ \fi}
\def \Nvir {\ifmmode N_{\rm  vir} \else $N_{\rm  vir}$ \fi}  
\def \Jvir {\ifmmode J_{\rm vir} \else $J_{\rm vir}$ \fi} 
\def \Jshell {\ifmmode J_{\rm shell} \else $J_{\rm shell}$ \fi}
\def \Evir {\ifmmode E_{\rm vir} \else $E_{\rm vir}$ \fi} 

\begin{abstract}
We allow simultaneously for a CDM--DE coupling and non--zero neutrino
masses and find that significant coupling and neutrino mass are
(slightly) statistically favoured in respect to a cosmology with no
coupling and negligible neutrino mass (our best fits are: $C \sim
1/2m_p$, $ m_\nu \sim 0.12\, $eV each flavor).  We assume DE to be a
self--interacting scalar field and use a standard Monte Carlo Markov
Chain approach.
%\vspace{1pc}
\end{abstract}

\maketitle

%\noindent
One of the main puzzles of cosmology is why a model as $\Lambda$CDM,
implying so many conceptual problems, apparently fits all linear data
\cite{bib1,bib2,bib3} in such unrivalled fashion.

The fine tuning paradox of $\Lambda$CDM is actually eased in dynamical
DE (dDE) cosmologies, where DE is a self--interacting scalar field,
facing no likelihood downgrade \cite{bib6}. The coincidence paradox is
also eased if an energy flow from CDM to dDE occurs (cDE cosmologies
\cite{coupling}). CDM--DE coupling, however, cuts the model
likelihood, as soon as an intensity suitable to attenuate coincidence
is approached.

The right physical cosmology could however include a further
ingredient, able to compensate coupling distorsions.  Here we shall
show a possible option of this kind: when we assume CDM--DE coupling
or a significant $\nu$ mass, we cause opposite spectral shifts
\cite{lavacca}. If we try to compensate them, the residual tiny
distorsions (slightly) favor coupling and $\nu$ mass, in respect to
dDE or $\Lambda$CDM.

In this paper we consider the self--interaction potentials, admitting
tracker solutions,
\begin{equation}
V(\phi) = \Lambda^{\alpha+4}/\phi^\alpha 
\end{equation}
or
\begin{equation}
V(\phi) = (\Lambda^{\alpha+4}/\phi^\alpha) \exp(4\pi\, \phi^2/m_p^2),
\end{equation}
(RP \cite{RP88} and SUGRA \cite{SUGRA}, respectively). Uncoupled RP
(SUGRA) yields a slowly (fastly) varying $w(a)$ state parameter.
Coupling is however an essential feature and modifies these behaviors,
mostly lowering $w(a)$ at low $z$, and boosting it up to +1, for $z
>\sim 10$.

For any choice of $\Lambda$ and $\alpha$ these cosmologies have a
precise DE density parameter $\Omega_{de}$. Here we take $\Lambda$ and
$\Omega_{de}$ as free parameters in flat cosmologies; the related
$\alpha$ value then follows.

In these scenarios, DE energy density and pressure read
\begin{equation}
\rho = \rho_k + V(\phi)~,~~~ p = \rho_k - V(\phi)~,
\end{equation}
{\rm with}~~
\begin{equation}
\rho_k = \dot \phi^2/2a^2~;
\label{rhop}
\end{equation}
dots indicating differentiation in respect to $\tau$, the background
metrics being
\begin{equation}
ds^2 = a^2(\tau) \left[ d\tau^2 - d\lambda^2 \right]
\label{metric}
\end{equation}
with
\begin{equation}
d\lambda^2 = dr^2 + r^2(d\theta^2 + cos^2 \theta\, d\phi^2)~.
\end{equation}
Until $\rho_k \gg V$, therefore, $w(a)=p/\rho$ approaches +1 and DE
density would rapidy dilute ($\rho \propto a^{-6}$), unless a feeding
from CDM occurs. When the $V \gg \rho_k$ regime is attained, then, the
state parameter approaches --1 and DE can account for cosmic
acceleration.

DE cannot couple to baryons, because of the equivalence principle
(see, {\it e.g.}  \cite{darmour1}). Constraints to CDM--DE
interactions, however, can only derive from cosmological data.

The simplest coupling is a linear one, formally obtainable through a
conformal transformation of Brans--Dicke theory (see, {\it e.g.},
\cite{brans}). The energy transfer from CDM to DE keeps then the
latter at a non--negligible density level even when $w(a) \sim
+1$. CDM density then declines slightly more rapidly than $a^{-3}$.
When $\phi$ approaches $m_p$ and $V(\phi) $ exceeds $\rho_k$, DE
dilution stops and DE eventually exceeds CDM density.

All tenable cosmological models comprise a 2--component dark sector
non interacting with baryons or radiation, so that $
T^{(c)~\mu}_{~~~~\nu;\mu} + T^{(de)~\mu}_{~~~\, ~~\nu;\mu} = 0 $
($T^{(c,de)}_{\mu\nu}:$ stress--energy tensors for CDM and DE, let
their traces read $T^{(c,de)}$). The assumption that CDM and DE are
separate leads to take $C \equiv 0$ in the relations
\begin{equation}
T^{(de)~\mu}_{~~~\, ~~\nu;\mu} = +C T^{(c)} \phi_{,\nu}~,~~
T^{(c)~\mu}_{~~~~\nu;\mu} =- C T^{(c)} \phi_{,\nu}~,
\end{equation}
describing the most general form of linear coupling (incidentally,
this shows why DE cannot couple to any component with vanishing
stress--energy tensor trace). Besides of $C$, we shall also use the
dimensionless coupling parameter $ \beta = (3/16\pi)^{1/2} m_p C~.  $

The results shown in this paper are based on our generalization of the
public program CAMB \cite{camb}, enabling it to study cDE models.
Likelihood distributions are then worked out by using CosmoMC
\cite{lewis:2002}.

\begin{figure}[htb]
\hskip -.5truecm
\includegraphics[height=6.8cm,width=8.cm]{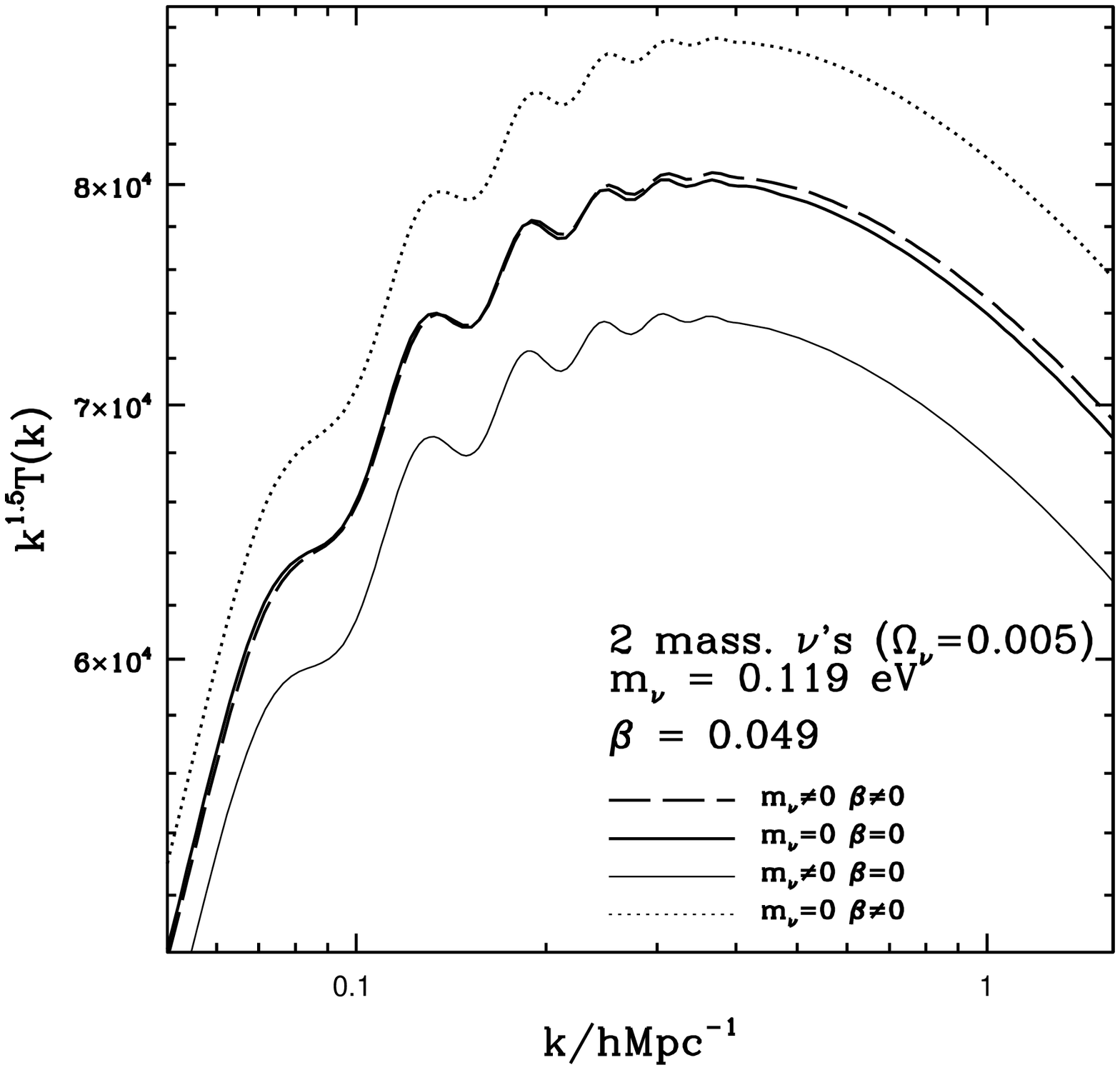}
%\hskip -2.9truecm
\includegraphics[height=8.5cm,width=8.cm]{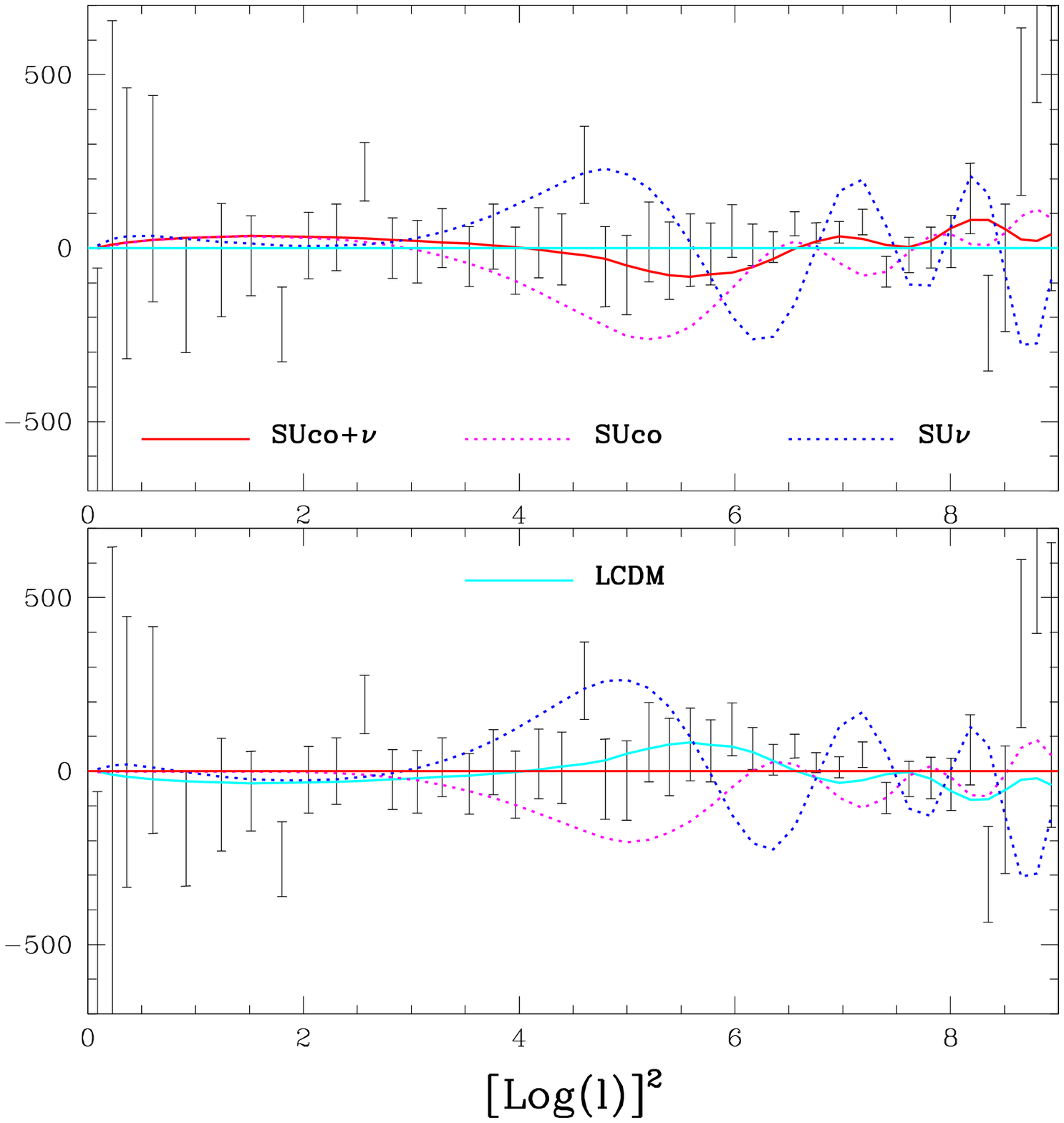}
\caption{{\bf Top panel:} Transfer functions (multiplied by $k^{1.5}$,
for graphic aims) in example cosmologies with/without coupling and
with/without 2 massive $\nu$'s (00/CM models). {\bf Bottom panel:}
Binned anisotropy spectral data, normalized to the best--fit
$\Lambda$CDM model (upper frame) or to the best fit SUGRA cDE model
including $\nu$--masses (lower frame).  The distorsions arising when
coupling or $\nu$--mass are separately considered are also shown.}
\label{fig1a}
%\vskip +.1truecm
\end{figure}

In Figure \ref{fig1a} we then illustrate the compensating effects
between coupling and $\nu$ mass. Details are provided in the caption.
%%%%%%%%%%%%%%%%%%%%%%%%%%%%%%%%%%%%%%%%%%%%%%%%%%%%%%%%%%%%%%%%%%%%
%%%%%%%%%%%%%%%%%%%%%%%%%%%%%%%%%%%%%%%%%%%%%%%%%%%%%%%%%%%%%%%%%%%%

\begin{figure}[htb]
\hskip -.9truecm
\begin{center}
\includegraphics[height=7.2cm,angle=-90]{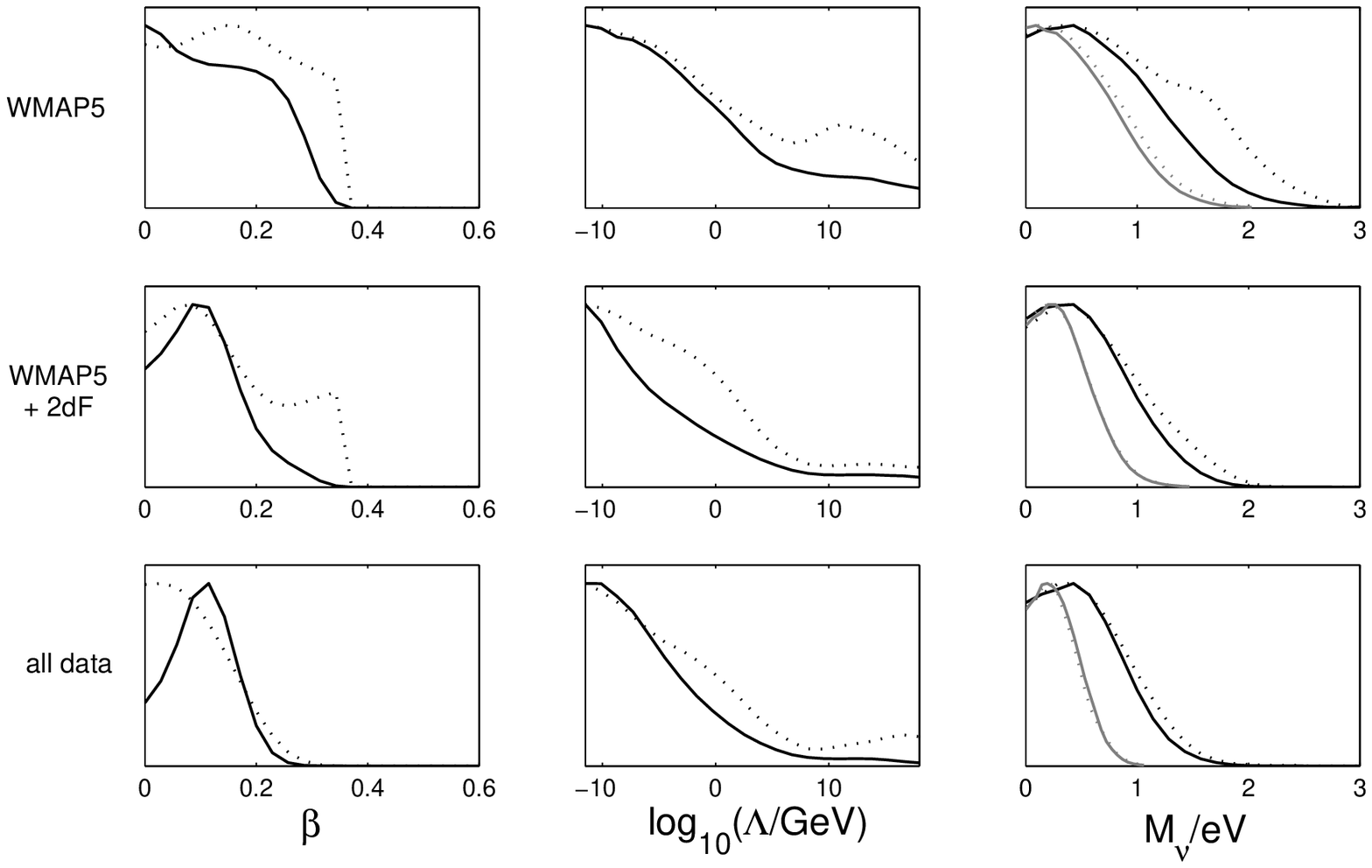}
\end{center}
%\vskip-.5truecm
\caption{Marginalized (solid line) and average (dotted line)
likelihood of cosmological parameters in SUGRA models.  Notice that
the $\beta$ signal appears when high-- and low--$z$ data are put
together, and is strengthened by SNIa data.  As a matter of fact,
coupling allows to lower the ``tension'' between $\Omega_c$ and
fluctuation amplitude detected from CMB and deep sample data. For
$M_{\nu}$ we also show the corresponding likelihood distributions
obtained in the case of a standard $\Lambda$CDM+$M_\nu$ model (gray
lines).}
\label{mnuall}
\end{figure}

\begin{figure}[htb]
%\vskip-.1truecm
\begin{center}
\includegraphics[height=7.2cm,angle=-90]{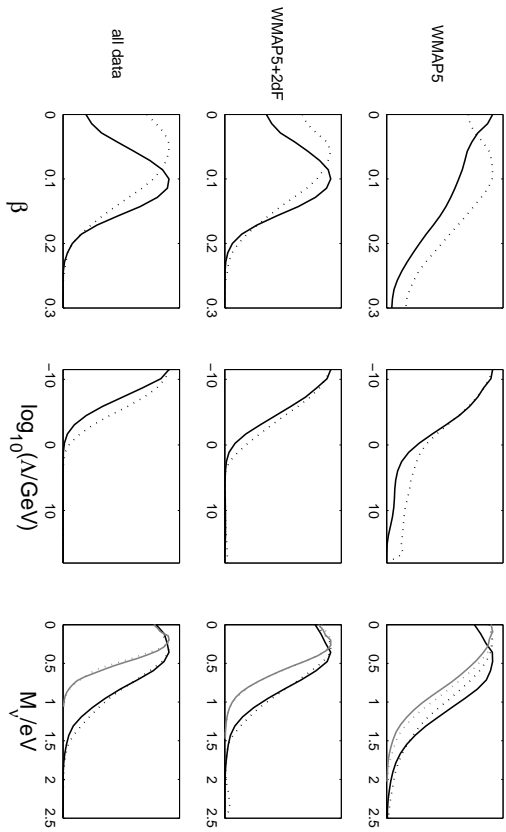}
\end{center}
%\vskip-.5truecm
\caption{As previous Figure, in RP models.}
\label{Rmnuall}
\end{figure}

When the fitting procedure is performed, we find most parameter values
in the same range as in dDE or $\Lambda$CDM cosmologies. The 
significant parameters in our approach are however the energy scale
$\Lambda$, the coupling intensity $\beta$ and the sum of
$\nu $ masses $M_\nu$. In Figures \ref{mnuall} and \ref{Rmnuall}
we provide one--dimensional likelihood distributions on these
parameters for SUGRA and RP cosmologies.
%%%%%%%%%%%%%%%%%%%%%%%%%%%%%%%%%%%%%%%%%%%%%%%%%%%%%%%%%%%%%%%%%%%%
%%%%%%%%%%%%%%%%%%%%%%%%%%%%%%%%%%%%%%%%%%%%%%%%%%%%%%%%%%%%%%%%%%%%
%%%%%%%%%%%%%%%%%%%%%%%%%%%%%%%%%%%%%%%%%%%%%%%%%%%%%%%%%%%%%%%%%%%%
%%%%%%%%%%%%%%%%%%%%%%%%%%%%%%%%%%%%%%%%%%%%%%%%%%%%%%%%%%%%%%%%%%%%
%%%%%%%%%%%%%%%%%%%%%%%%%%%%%%%%%%%%%%%%%%%%%%%%%%%%%%%%%%%%%%%%%%%%
\begin{figure}[htb]
\begin{center}
\includegraphics[height=5.5cm,angle=0]{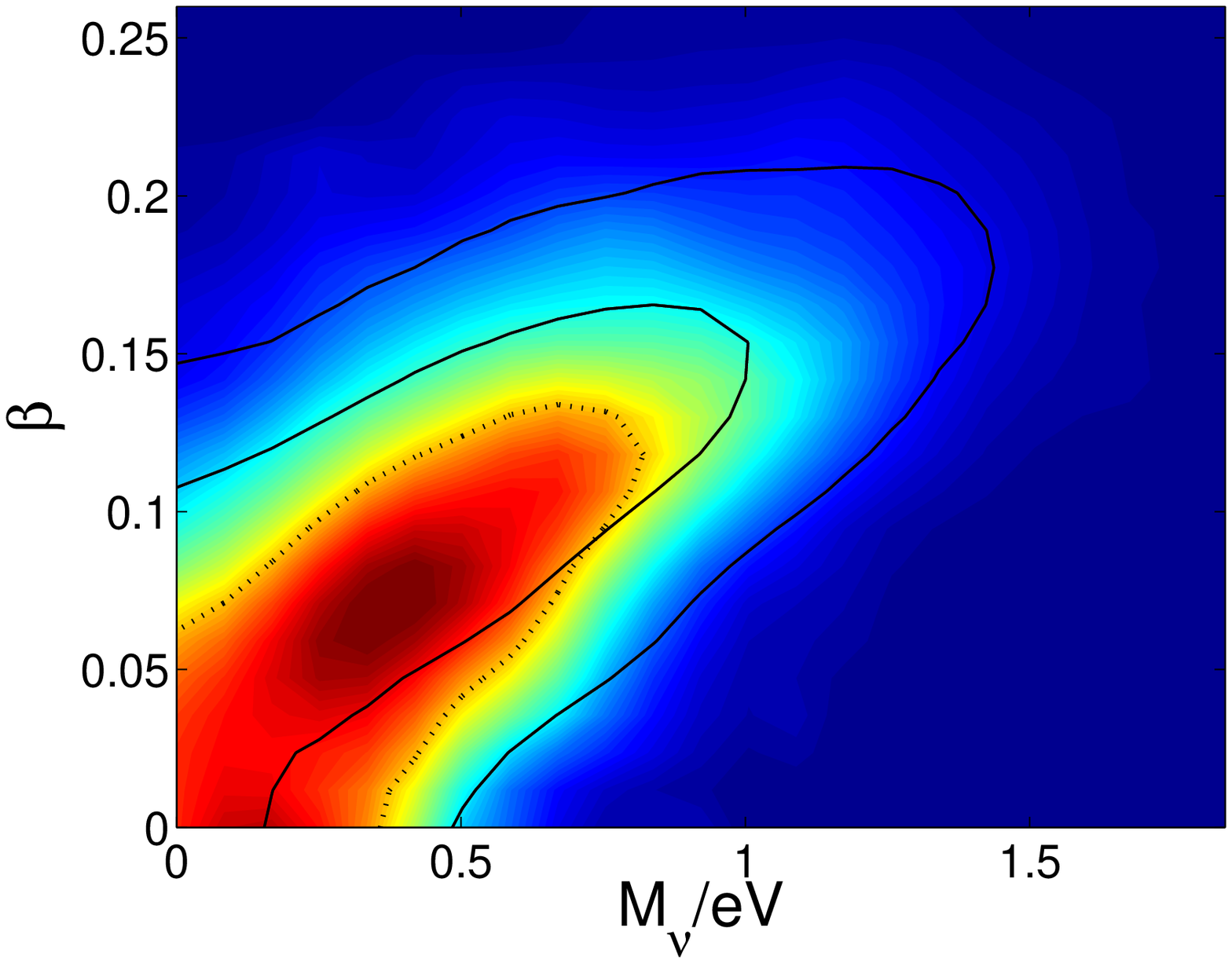}
\vskip.5truecm
\includegraphics[height=5.5cm,angle=0]{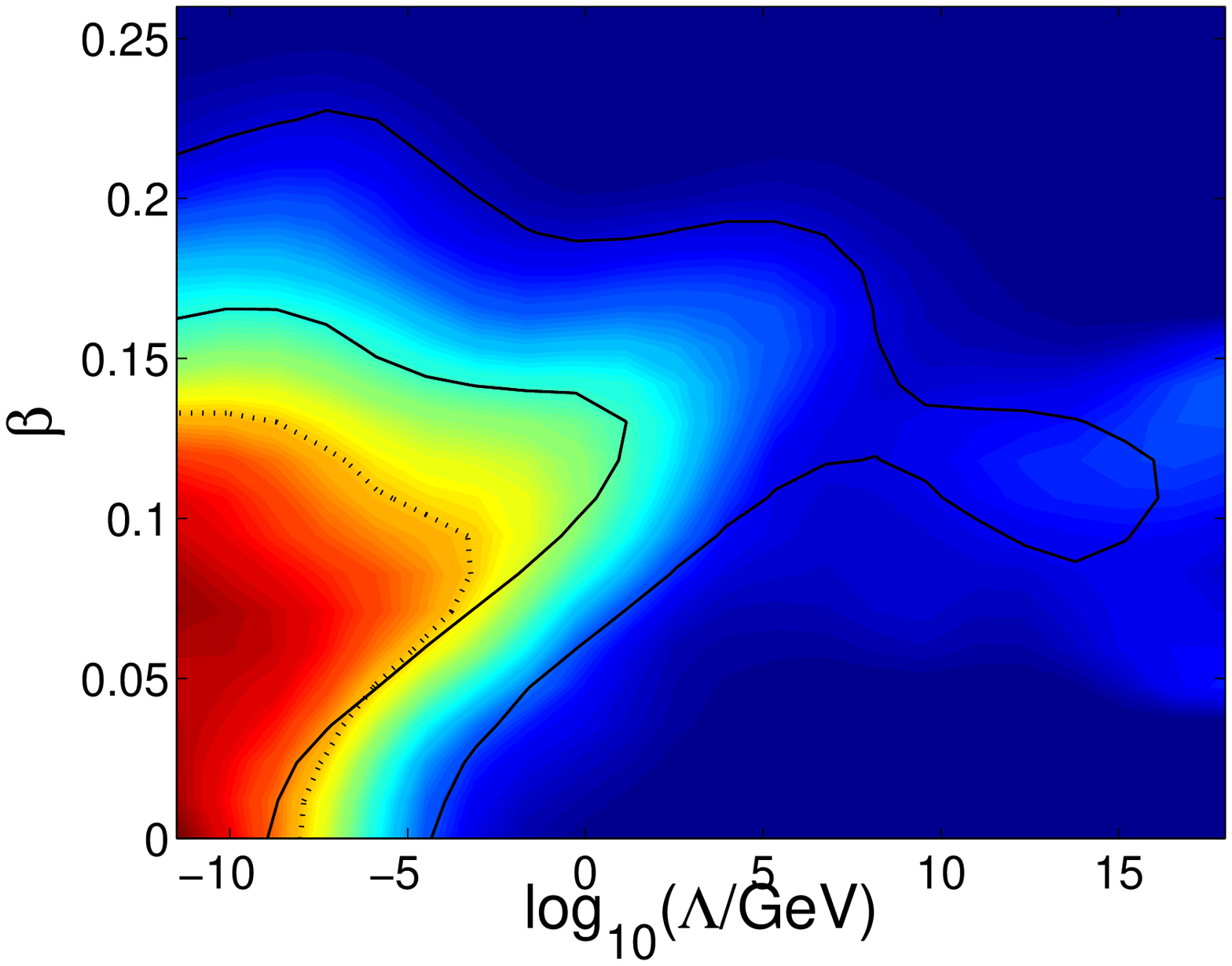}
\end{center}
\caption{Two parameter contours for the SUGRA model. Solid lines are
1-- and 2--$\sigma$ limits for marginalized likelihood.  Colors refer
to average likelihood, and the 50\% likelihood contour from the
average likelihood is indicated by the dotted line.}
\label{2DS}
\end{figure}
%%%%%%%%%%%%%%%%%%%%%%%%%%%%%%%%%%%%%%%%%%%%%%%%%%%%%%%%%%%%%%%%%%%%
%%%%%%%%%%%%%%%%%%%%%%%%%%%%%%%%%%%%%%%%%%%%%%%%%%%%%%%%%%%%%%%%%%%%
\begin{figure}[htb]
\begin{center}
\includegraphics[height=5.5cm,angle=0]{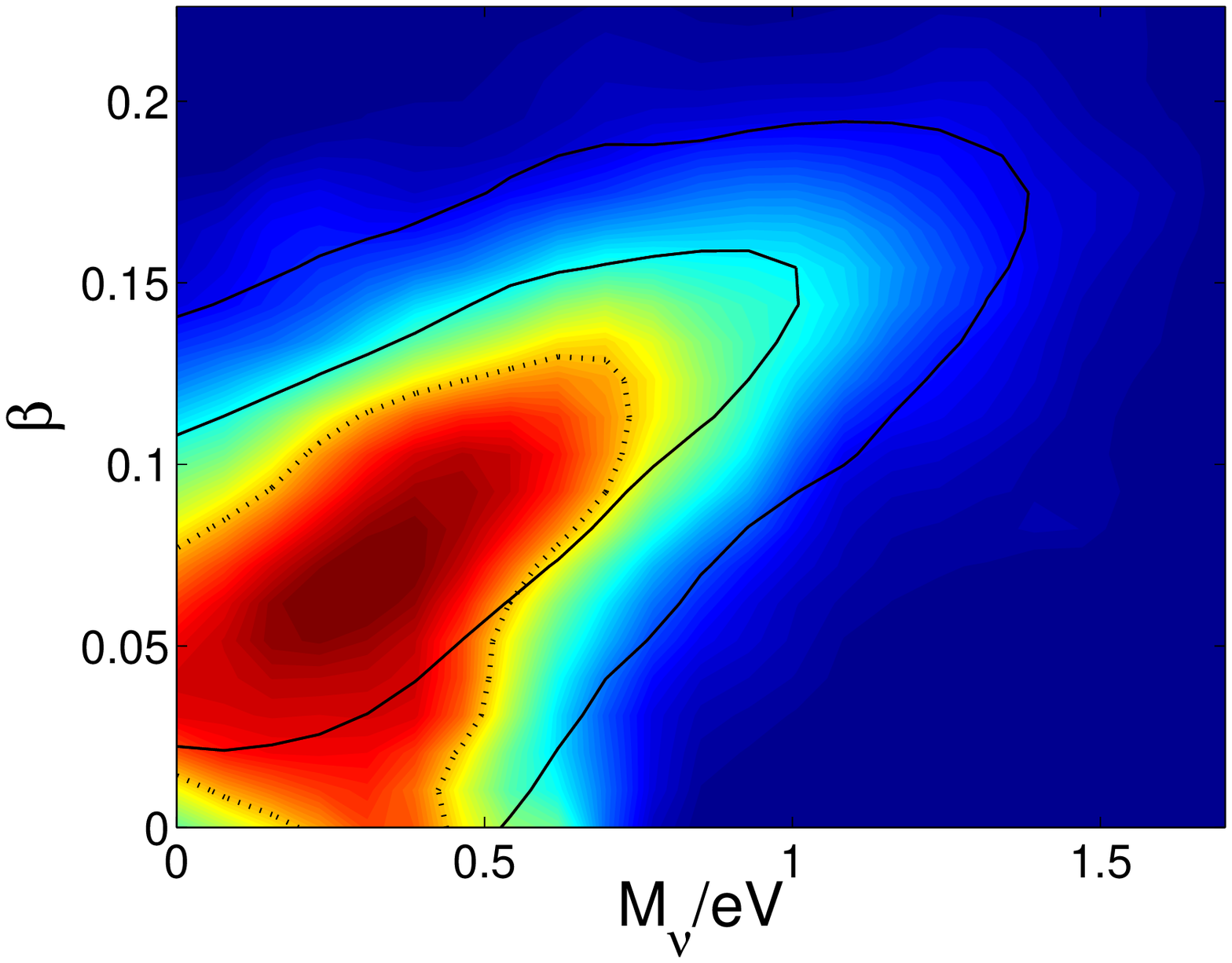}
\vskip.5truecm
\includegraphics[height=5.5cm,angle=0]{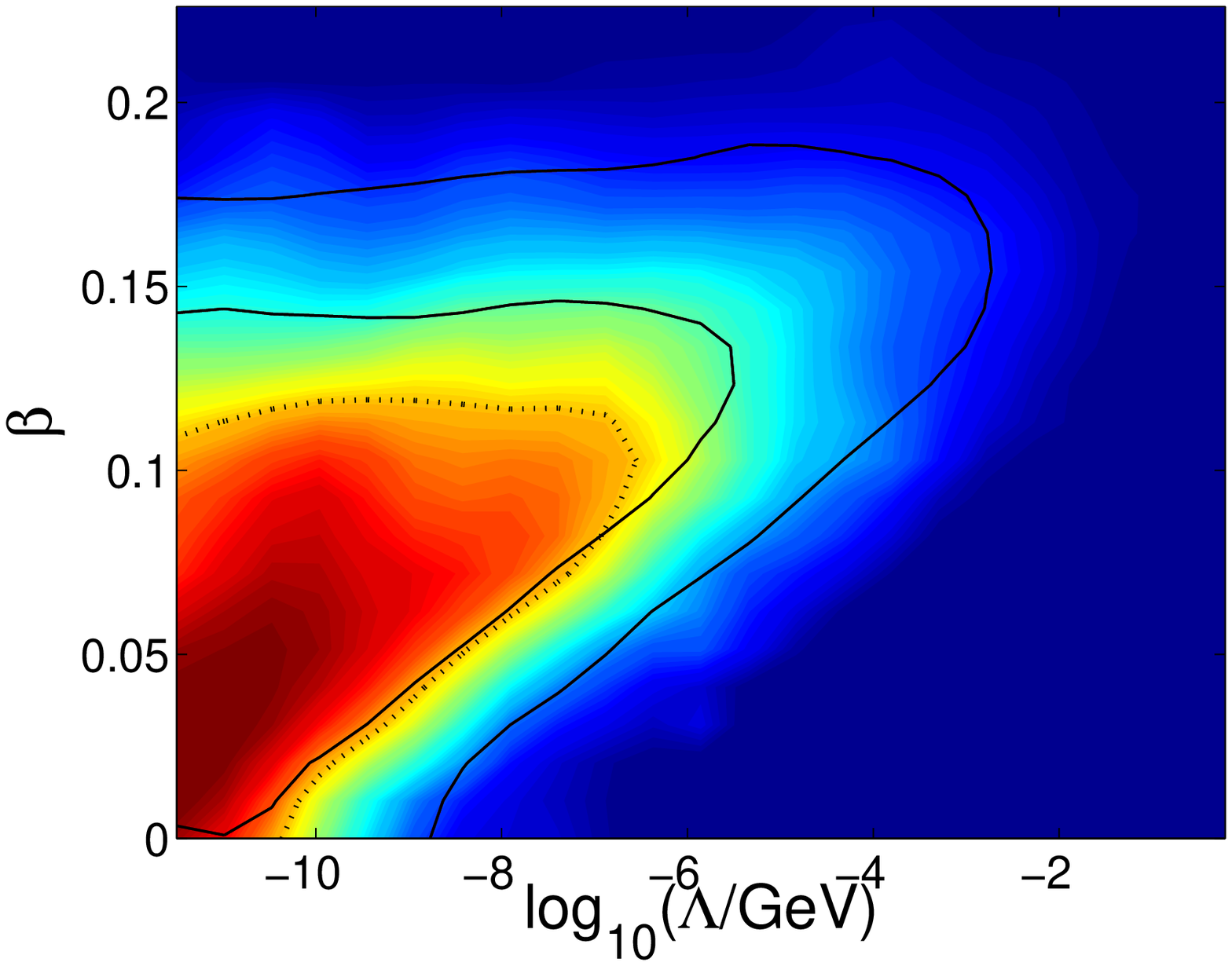}
\end{center}
\caption{As previous Figure, for a RP potential.}
\label{2DR}
\end{figure}
%%%%%%%%%%%%%%%%%%%%%%%%%%%%%%%%%%%%%%%%%%%%%%%%%%%%%%%%%%%%%%%%%%%%
%%%%%%%%%%%%%%%%%%%%%%%%%%%%%%%%%%%%%%%%%%%%%%%%%%%%%%%%%%%%%%%%%%%%
\begin{figure}[htb]
\begin{center}
\includegraphics[height=6.cm,width=8.2truecm,angle=0]{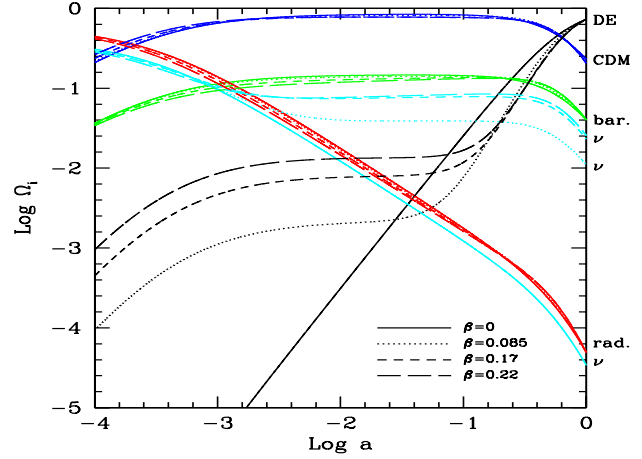}
\end{center}
\vskip -.3truecm
\caption{Evolution of density parameters in a SUGRA model with
coupling and $\nu$ mass. Colors refer to different components, as
specified in the frame; lines to the different models: $M_\nu=0,
~\beta=0$ (continuous line); $M_\nu=0.5~$eV, $\beta = 0.085$ (dotted);
$M_\nu=1.1~$eV, $\beta = 0.17$ (short dashed); $M_\nu=1.2~$eV, $\beta
= 0.22$ (long dashed). }
\label{RPom}
\end{figure}
%%%%%%%%%%%%%%%%%%%%%%%%%%%%%%%%%%%%%%%%%%%%%%%%%%%%%%%%%%%%%%%%%%%%
%%%%%%%%%%%%%%%%%%%%%%%%%%%%%%%%%%%%%%%%%%%%%%%%%%%%%%%%%%%%%%%%%%%%
\begin{figure}[h!]
\begin{center}
\includegraphics[width=7.cm,angle=0]{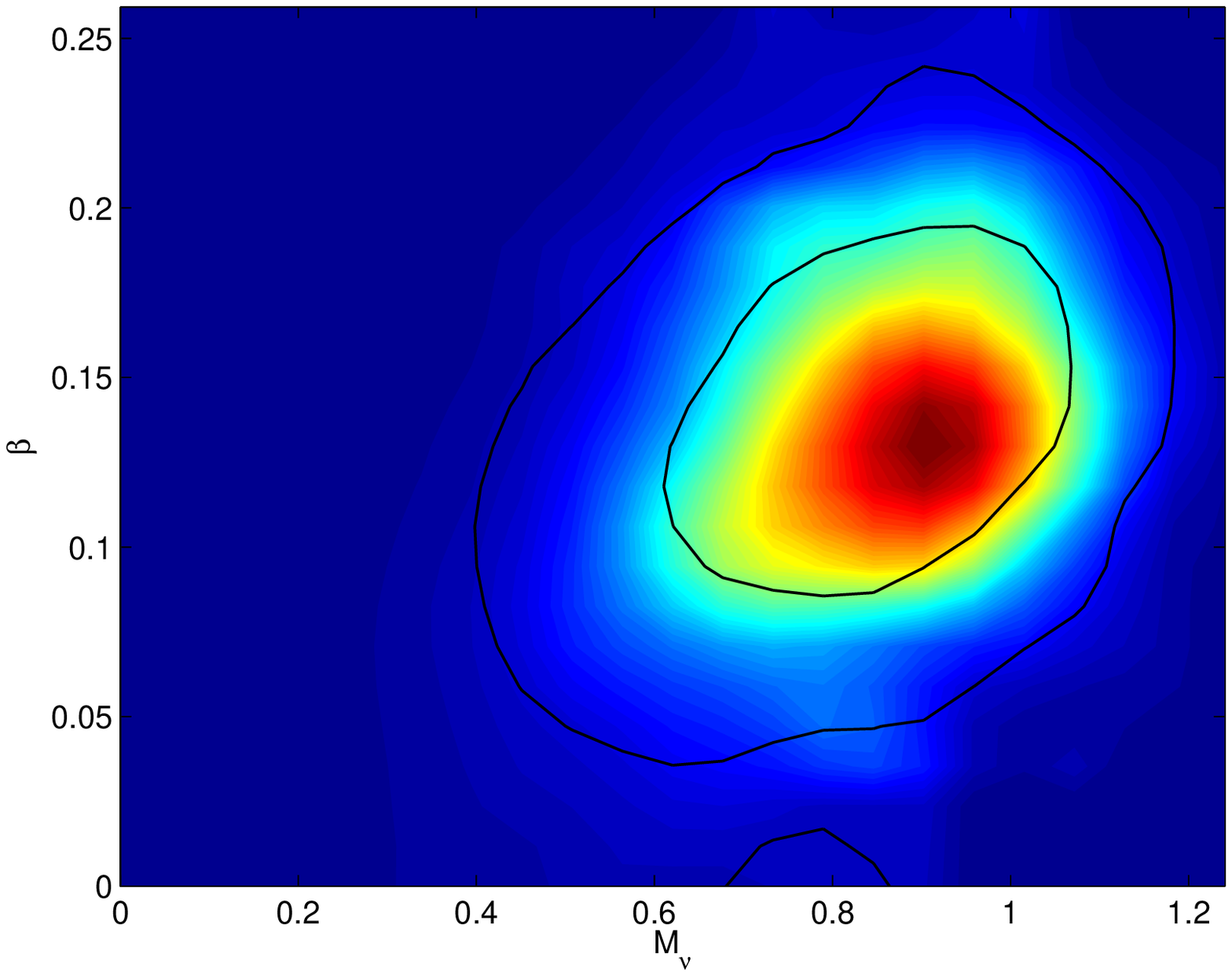}
\end{center}
%\vskip -.9truecm
\caption{Detection of CDM--DE coupling, following an hypothetical
determination of an electron neutrino mass $m_\nu \simeq 0.3~$eV, by
the experiment KATRIN, yielding $M_\nu \simeq 0.9~$eV.  Here we show
the likelihood distribution, with such a prior on $M_\nu$. Colors and
lines convey the same indications as in Figures 4 and 5.  }
\label{katrin}
\end{figure}
%%%%%%%%%%%%%%%%%%%%%%%%%%%%%%%%%%%%%%%%%%%%%%%%%%%%%%%%%%%%%%%%%%%%

A basic information is however the correlation between likelihood
distributions. These are shown in Figures \ref{2DS} and \ref{2DR}
again for SUGRA and RP cosmologies, respectively.

These Figures, as well as one--dimensional plots, clearly exhibit
maxima, both for average and marginalized likelihood, for
significantly non--zero coupling and $\nu$--masses. Although their
statististical significance is not enough to indicate any
``detection'' level, the indication is impressive. Furthermore, a
stronger signal, with the present observational sensitivity, would be
impossible.

Let us however outline that this work aimed at finding how far one
could go from $\Lambda$CDM, adding non--zero coupling and $\nu$--mass,
without facing a likelihood degrade. It came then as an unexpected
bonus that likelihood does not peak on the 0--0 option.

The allowed $\beta$ values open the possibility of a critically
modified DE behavior. Figure \ref{RPom} shows the scale dependence of
the cosmic components for various $\beta$--$M_\nu$ pairs.

Let us then outline that the $M_\nu$ values allowed here approach the
$\nu$--mass detection area in the forthcoming tritium decay experiment
KATRIN \cite{katrin}. 
Should particle data lead to an external prior on $M_\nu$, the strong
degeneracy between the coupling parameter $\beta$ and the neutrino
mass $M_\nu$ is broken, and new insight into the DE nature is gained
\cite{kristiansen09}. In Figure \ref{katrin} we show how a neutrino
mass determination symultaneously implies an almost model independent
CDM--DE coupling detection. This would be a revival of mixed DM models
\cite{mix}, in the form of Mildly Mixed Coupled (MMC) cosmologies.

%\section*{References}

\newcommand{\Nature}{{\it Nature\/} }
\newcommand{\ApJ}{{\it Astrophys. J.\/} }
\newcommand{\ApJS}{{\it Astrophys. J. Suppl.\/} }
\newcommand{\MNRAS}{{\it Mon. Not. R. Astron. Soc.\/} }
\newcommand{\PhRv}{{\it Phys. Rev.\/} }
\newcommand{\PhL}{{\it Phys. Lett.\/} }
\newcommand{\JCAP}{{\it J. Cosmol. Astropart. Phys.\/} }
\newcommand{\AeA}{{\it Astronom. Astrophys.\/} }
\newcommand{\etall}{{\it et al.\/} }
\newcommand{\arXiv}{{\it Preprint\/} }

\end{document}